\documentclass{article}

\usepackage{arxiv}

\usepackage[utf8]{inputenc} 
\usepackage[T1]{fontenc}    
\usepackage{hyperref}       
\usepackage{url}            
\usepackage{booktabs}       
\usepackage{amsfonts}       
\usepackage{nicefrac}       
\usepackage{microtype}      
\usepackage{lipsum}
\usepackage{graphicx}

\usepackage{xcolor}         
\usepackage{subfiles}
\usepackage{algorithm}
\usepackage{algorithmicx}
\usepackage{amsmath}
\usepackage{amsthm}
\usepackage{algorithm}
\usepackage{algorithmicx}
\usepackage{subfigure}
\usepackage[noend]{algpseudocode}
\usepackage{array}
\usepackage{balance}

\newtheorem{theorem}{Theorem}

\newtheorem{Defs}{Definition}

\title{Enhanced Privacy Bound for Shuffle Model \\ with Personalized Privacy}

\author{
 Yixuan Liu \\
  Renmin University of China \\
  \texttt{liuyixuan@ruc.edu.cn} \\
   \And
 Yuhan Liu \\
  Renmin University of China\\
  \texttt{liuyh2019@ruc.edu.cn} \\
  \And
 Li Xiong \\
  Emory University\\
  \texttt{lxiong@emory.edu} \\
  \AND
 Yujie Gu \\
  Kyushu University \\
  \texttt{gu@inf.kyushu-u.ac.jp} \\
  \And
 Hong Chen \\
  Renmin University of China \\
  \texttt{chong@ruc.edu.cn} \\
}
\begin{document}
\maketitle

\begin{abstract}
The shuffle model of Differential Privacy (DP) is an enhanced privacy protocol which introduces an intermediate trusted server between local users and a central data curator. It significantly amplifies the central DP guarantee by anonymizing and shuffling the local randomized data.
Yet, deriving a tight privacy bound is challenging due to its complicated randomization protocol.  
While most existing work are focused on unified local privacy settings, this work focuses on deriving the central privacy bound for a more practical setting  where personalized local privacy is required by each user. To bound the privacy after shuffling, we first need to capture the probability of each user generating clones of the neighboring data points. Second, we need to quantify the indistinguishability between two distributions of the number of clones on neighboring datasets.
Existing works either inaccurately capture the probability, or underestimate the indistinguishability between neighboring datasets.
Motivated by this, we develop a more precise analysis, which yields a general and tighter bound for arbitrary DP mechanisms. Firstly, we derive the clone-generating probability by hypothesis testing 
, which leads to a more accurate characterization of the probability.
Secondly, we analyze the indistinguishability in the context of $f$-DP, where the convexity of the distributions is leveraged to achieve a tighter privacy bound.
Theoretical and numerical results demonstrate that our bound remarkably outperforms the existing results in the literature.

\end{abstract}



\section{Introduction}

The shuffle model \cite{bittau2017prochlo} with Differential Privacy (DP) \cite{dwork2006calibrating} is an advanced privacy protection protocol for distributed data analysis \cite{girgis2021shuffled, liu2021privacy, chen2024privacy, scott2022aggregation}. An intermediate trusted server \emph{shuffler} is introduced between local randomizer \cite{erlingsson2014rappor} and central analyzer \cite{dwork2014algorithmic}. By  permuting locally randomized data before sending to the central analyzer, the shuffler brings extra randomness with a privacy amplification effect, i.e., central privacy guarantee after shuffling significantly stronger than the original local privacy is achieved by perturbation. 

\begin{figure}
\setlength{\abovecaptionskip}{5pt}%
\setlength{\belowcaptionskip}{5pt}%
  \centering
  \includegraphics[width=0.65\linewidth, trim=20 10 20 10,clip]{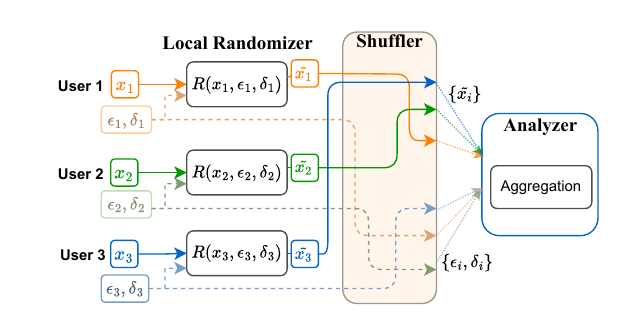}
  \caption{Procedure of shuffle model with personalized privacy. Each user data $x_i$ is randomized locally. Privacy parameters $(\epsilon_i, \delta_i)$ and perturbed $\tilde{x}_i$ are shuffled. Analyzer aggregates $\tilde{x}_i$ for further statistics or model training.}
  \label{pics_framework}
\end{figure}

Many efforts have been put on converting the randomness to a formal privacy guarantee
\cite{balle2019privacy,erlingsson2019amplification,feldman2022hiding, feldman2023stronger, girgis2024multi}. While most studies achieve privacy bound by assuming a unified privacy level for all users, this work focuses on a more practical but less studied setting with personalized privacy, where users have different privacy levels on local data points due to different policies or privacy preferences \cite{liu2021projected,liu2023echo,liu2023personalized,liu2024cross}. (Fig. \ref{pics_framework} shows the personalized setting where local data point $x_i$ is associated with a personalized privacy level $\epsilon_i$, $\delta_i$) 

A classic privacy analysis for shuffle model 
amplifies the privacy by leveraging the confounding effect of clones of neighboring data points generated by each user \cite{feldman2022hiding, feldman2023stronger}. Specifically, for any neighboring datasets that differ by $x_1$, the noisy data point from each user could generate a clone of randomized $x_1$ with certain probability $p$. The clones together help to hide the existence of $x_1$; then the difference of the number of clones on neighboring datasets is estimated for final privacy bounds.

However, driving the probability $p$ and the difference of number of clones is challenging, especially under Personalized Local Differential Privacy (PLDP) setting. 
Approximating $p$ with conventional way that reduces any DP randomized mechanism to the worst-case random response leads to inaccurate results. Additionally, various privacy parameters exaggerate the complexity of the overall distributions of number of clones. Existing works \cite{liu2023echo, chen2024generalized} approximating it by central limit theorem causes relaxations on privacy bound, especially when the number of users is not large enough.

Motivated by this, we develop a more precise analysis on privacy amplification of shuffle model under both pure- and approximate-PLDP
for arbitrary local randomizers. 

Firstly, we quantify different $p$ contributed by each user with personalized privacy parameters in a more accurate manner. In specific, $p$ is derived by conducting hypothesis testing on the distribution of current noisy data point and the distribution of noisy neighboring data point. By computing the hypothesis testing error, we accurately identify the probability of any data points being wrongly-recognized as  $x_1$. Our method allows computing $p$ on heterogeneous privacy parameters and arbitrary DP local randomizer.

Secondly, we analyze the indistinguishability between two overall distributions of the number of clones in the context of  $f$-DP \cite{dong2022gaussian}. We depict the overall distributions by Multi-Bernoulli and Binomial distribution. Then inspired by \cite{wang2024unified}, the convexity of the distribution is further exploited to closely characterize the properties of the overall distributions, thus leading to tighter upper bound on the privacy after shuffling.

Our main contributions are summarized as follows: 
\begin{itemize}
    \item We provide a more precise analysis for privacy amplification effect on shuffle model for personalized privacy. Confounding effect of individuals and overall distributions are characterized by analytical expression, which leads to a tighter privacy bound.
    \item Our work offers a general method to quantify confounding effect of PLDP with hypothesis testing, which enables our analysis to address arbitrary locally differentially private mechanisms and heterogeneous privacy parameters. 
    \item We verify the proposed analysis with numerical results, which demonstrates that our privacy bound significantly exceeds the SOTAs on both pure- and approximate-PLDP.
\end{itemize}

\section{Preliminaries}
\subsection{Central and Local Differential Privacy}
Differential privacy (DP) \cite{dwork2014algorithmic} provides a rigorous privacy guarantee for raw data by introducing random noises to the computation process. 
The notion is typically applied in a central setting where a trusted server can access the raw data. For the settings without trusted server, local differential privacy (LDP) \cite{erlingsson2014rappor} is proposed. LDP is capable of providing a stronger privacy guarantee than DP, as it protects data against stronger adversaries who have access to every (perturbed) data point in the dataset. Therefore, it is suitable for distributed data collecting or publishing \cite{cormode2018privacy, wang2019collecting, liu2022collecting, liu2024edge}. Yet, LDP also suffers from a dissatisfying data utility due to a large amount of noise injection.
\begin{Defs}[Differential Privacy]
	For any $\epsilon, \delta \geq 0$, a randomized algorithm $R: \mathcal{D} \rightarrow \mathcal{Z}$ is $(\epsilon, \delta)$-DP if for any neighboring datasets $D, D' \in \mathcal{D}$ and any subsets $S \subseteq \mathcal{Z}$,
	$\Pr[R(D) \in S] \leq e^{\epsilon}\Pr[R(D') \in S] + \delta$.
\end{Defs}
\begin{Defs}[Local Differential Privacy]
	For any $\epsilon, \delta \geq 0$, a randomizer $R: \mathcal{D} \rightarrow \mathcal{Z}$  is $(\epsilon, \delta)$-LDP if $\ \forall x, x' \in \mathcal{D}$ and $\forall z \in \mathcal{Z}$,
	$\Pr[R(x) = z] \leq e^{\epsilon}\Pr[R(x') = z ] + \delta$.
\end{Defs}
$\epsilon$ denotes the privacy level, the lower the stronger privacy. $\delta$ denotes the failure probability of the randomizer. $\delta=0$ is pure-LDP, and $\delta>0$ is approximate-LDP.


\subsection{Shuffle-based Privacy}
Shuffle model \cite{bittau2017prochlo} is proposed to strengthen central privacy while preserving local user privacy. 
Given dataset $D$, each $x_i \in D$ owned by user $i$ is perturbed locally by a randomizer $R$ to ensure $(\epsilon_i^l, \delta_i^l)$-LDP and sent to shuffler. 
Shuffler $S$, a trusted third party, permutes all data points and sends them to an untrusted analyzer $A$ for further computation. Based on the anonymity from shuffling, existing works obtain a strong privacy amplification effect. Most of works \cite{erlingsson2019amplification, balle2019privacy,feldman2022hiding, feldman2023stronger, wang2024unified, girgis2024multi} focus on unified local privacy setting, \cite{feldman2023stronger} improves privacy bound by generating clones from neighbor data points. \cite{wang2024unified} applies $f$-DP and achieves a tighter bound under unified LDP. As a more common and practical setting, 
some works \cite{liu2023echo, chen2024generalized} focus on personalized settings, while leaving a loose privacy bound due to reduction or approximation.

\section{Privacy Analysis}
In this section, we first introduce the confounding effect, which captures the randomness introduced by shuffler and serves as the
foundation of amplification effect analysis. 
Then we provide an analytical expression of confounding effect with hypothesis testing, 
which yields a precise description  
and results in a stronger amplification effect.
At last, we develop our analysis in the context of $f$-DP. By exploiting the convexity of the mixed distribution generated by the shuffler, we further derive a tighter bound for both pure- and approximate-PLDP.

\subsection{Confounding Effect $p$}
We consider neighboring data points $x_1^0$ and $x_1^1 \in D$. As noted in \cite{feldman2022hiding}, after perturbing and shuffling each data point, the output of randomizer on each data point could be seen as samples from the output distribution of randomizer on $x_1^0$ or $x_1^1$ with certain probability.  And each local randomizer $R(x, \epsilon):D \rightarrow Z$ can also be 
represented as: $R(x_1^0)=(1-p) Q(x_1^0) + p Q(x_1^1)$ and $R(x_1^1)= p Q(x_1^0) + (1-p) Q(x_1^0)$, where 
$Q:\{x_1^0,x_1^1\} \rightarrow S$ is a randomized algorithm. Hence the following decomposition is given by \cite{feldman2022hiding}:
\begin{align}
R(x_1^0) = e^\epsilon p Q(x_1^0) + p Q(x_1^1), \quad
R(x_1^1) = p Q(x_1^0) + e^\epsilon p Q(x_1^1) \label{eq_FV_2}\\ 
\forall i\in [2,n], \quad R(x_i) = p R(x_1^0) + p R(x_1^1) + (1-2p) LO_i
\label{eq_FV_3}
\end{align}
where $LO_i$ is the leftover distributions. The decomposition above suggests that each output from $R(x_i)$ could be wrongly recognized as coming from $x_1^0$ or $x_1^1$ with probability $p$. 
In other word, $p$ is the \emph{confounding effect} of $R(x_i)$ on $x_1^b$, where $b=0$ or $1$, and stronger privacy is achieved with a larger $p$. Existing works derive $p$ by reducing the LDP mechanism to random response \cite{kairouz2015composition}, which underestimates the confounding effect of most LDP mechanisms. 

In this work, we achieve a precise $p$. By conducting hypothesis testing on distributions $R(x_i)$ and $R(x_1^b)$ \footnote{For convenience, we 
use the simplified notation $R(x_i)$ instead of $R(x_i, \epsilon_i, \delta_i)$
when it is clear from the context, as $(\epsilon_i, \delta_i)$ is always binding with $x_i$.}, the type \uppercase\expandafter{\romannumeral1} error captures the probability of 
wrongly recognizing
the output of $R(x_i)$ as an output of $R(x_1^b)$, which is exactly $p$. For clarity, we derive the value of $p$ for the neighboring data point $x_1^b$ and the rest data point $x_i$ in Section \ref{sec_unified_eps} and \ref{sec_personal_eps} respectively.

\subsection{Quantifying $p$ with Hypothesis Testing}
\subsubsection{Hypothesis Testing on Neighboring Data Point $x_1^b$}
\label{sec_unified_eps}
In this section, we demonstrate our  hypothesis testing based approach for deriving $p$ at $x_1^b$, where the confounding of $R(x_1^b)$ only depends on the the privacy budget $(\epsilon_1, \delta_1)$.
Given a random output $Z$ from $R(x_1^b)$, we set the hypothesis testing as follows: \\
\centerline{$ H_0$: $Z$ came from $x_1^0$, \qquad $H_1$: $Z$ came from $x_1^1$.} \\
Then we conduct likelihood ratio test by examining the ratio between probability $p_1^0 = \Pr[R(x_1^0)=Z]$ and $p_1^1 = \Pr[R(x_1^1)=Z]$, and reject $H_0$ when $p_1^0 / p_1^1 < 1$. The rejection region is defined as 
$$S=\{z | \Pr[R(x_1^0)=z] < \Pr[R(x_1^1)=z] \}.$$
According to Neyman–Pearson lemma \cite{lehmann2006theory}, likelihood ratio test is the most powerful way to distinguish two distributions. Hence with such $S$, we achieve the lower bound of $p$.
As for approximate-DP, the privacy protection fails when outputs $z \in T_\delta$ where
$$T_\delta = \{z | \Pr[R(x)=z] < -\delta/2 ~\text{or}~ \Pr[R(x)=z] > 1-\delta/2\}.$$
After removing the failure set $T_\delta$, the $p$ is lower bounded by
$$\Pr[R(x_1^0) \in S\backslash T_{\delta_1}^0] = \Pr[R(x_1^1) \in \bar{S}\backslash T_{\delta_1}^0] = p_1.$$
where $\bar{S}$ is the complement of $S$, $T_{\delta_1}^0$ denotes the failure set on $x_1^0$ with $\delta_1$, $T_{\delta_1}^1$ is on $x_1^1$ with $\delta_1$.
Then Eq.\eqref{eq_FV_2} is rewritten as Eq. \eqref{eq_hp_2}. By further considering the distribution of concrete DP mechanisms, we are able to achieve exact expression of $p$. 
\begin{align}
R(x_1^0) = (\!1\!-\!p_1\!) R(x_1^0) \!+\! p_1 R(x_1^1),\quad 
R(x_1^1) = p_1 R(x_1^0) \!+\! (\!1\!-\!p_1\!) R(x_1^1) \label{eq_hp_2}
\end{align}


\begin{figure} 
  \centering
  \includegraphics[width=0.4\linewidth, trim=6 8 6 8,clip]{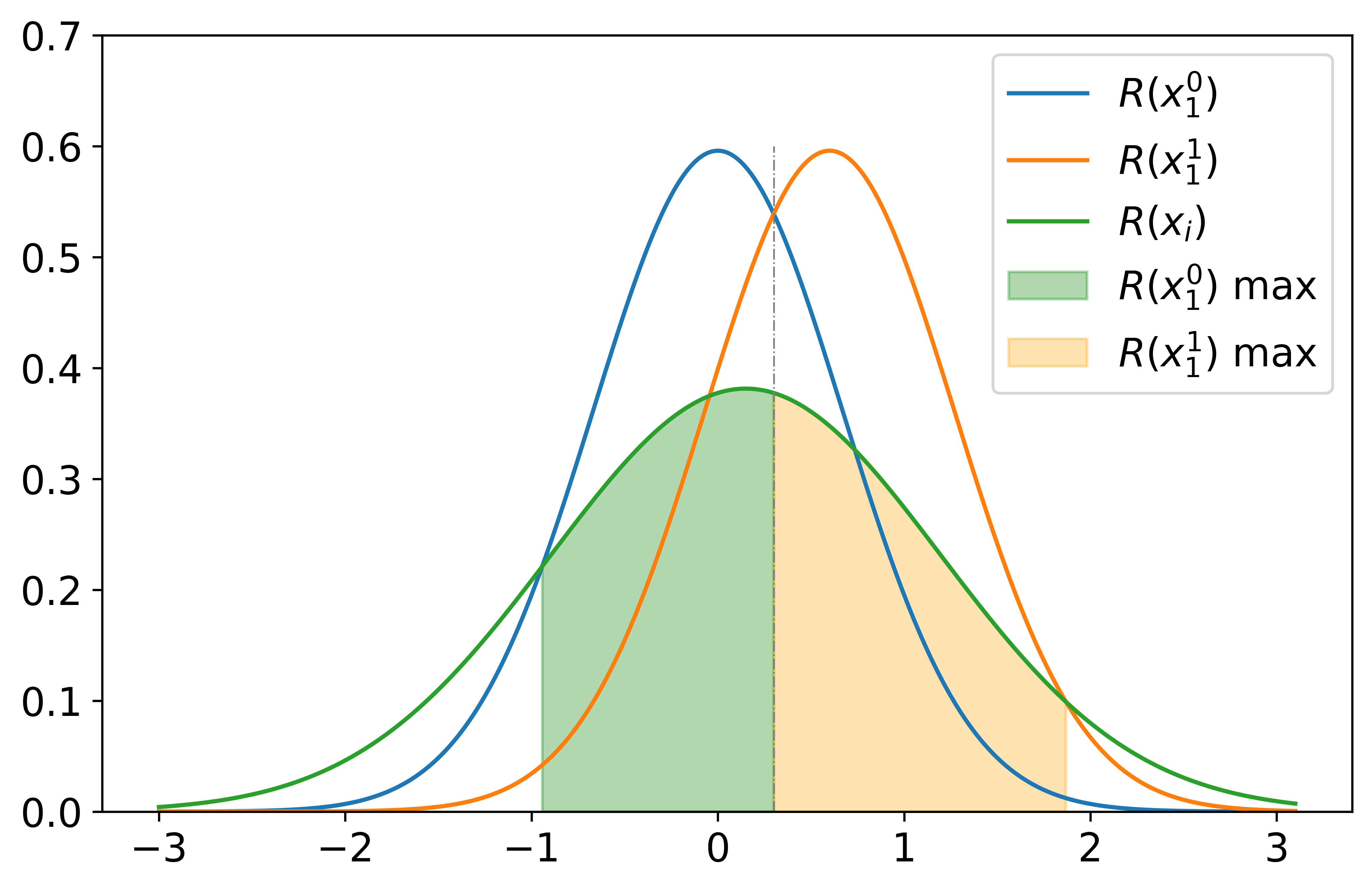}
  \caption{Green area represents $\Pr[R(x_i) \in U_0]$, output of $R(x_i)$ is wrongly recognized from $x_1^0$; Yellow area represents $\Pr[R(x_i) \in U_1]$, output of $R(x_i)$ is mistaken from $x_1^1$.}
\label{pics_dist}
\end{figure}

\subsubsection{Hypothesis Testing on Rest Data Points $x_i$}
\label{sec_personal_eps}
We then extend the method to $x_i$ for $i\in[2,n]$. The main difference lies in the confounding effect that involves heterogeneous privacy parameters $(\epsilon_1, \delta_1)$ and $(\epsilon_i, \delta_i)$ now.
Given a random output $Z$ of $R(x_i, \epsilon_i, \delta_i)$ and $R(x_1^b, \epsilon_1, \delta_1)$, we set hypothesis testing: \\
\centerline{$ H_0$: $Z$ came from $x_i$, \qquad $H_1$: $Z$ came from $x_1^b$.} \\
Noticing that $H_1$ indicates $Z$ came from $x_1^0$ or $x_1^1$, with likelihood ratio test, we set the rejection region as 
\begin{align*}
    U= \{ z| \Pr[R(x_i)=z] <  \max(\Pr[R(x_1^0)=z], \Pr[R(x_1^1)=z ]) \}
\end{align*}
Therefore, with $\Pr[R(x_i) \in U]$ null hypothesis is true but rejected, i.e., $R(x_i)$ is wrongly recognized as $R(x_1^b)$. 
$U$ could be further partitioned into two subsets $U_0$ and $U_1$:
\begin{align*}
    U_0=\!\{ z| \Pr[R(x_i)\!=\!z]\!<\!\Pr[R(x_1^0)\!=\!z] ~\text{and}~ \Pr[R(x_1^1)\!=\!z]\!< \!\Pr[R(x_1^0)\!=\!z] \} \\
    U_1=\!\{ z| \Pr[R(x_i)\!=\!z]\!<\!\Pr[R(x_1^1)\!=\!z] ~\text{and}~ \Pr[R(x_1^0)\!=\!z]\!<\!\Pr[R(x_1^1)\!=\!z]\}
\end{align*}
where $U_0 \cup U_1=U$ (Cf. Fig. \ref{pics_dist}). Similar with Section \ref{sec_unified_eps}, the failure set due to $\delta_i$ is removed from $U$. Accordingly, the probabilities of type \uppercase\expandafter{\romannumeral1} error on $x_1^0$ and $x_1^1$ are defined as:
$$p_i^0 = \Pr[R(x_i) \in U_0 \backslash T_{\delta_i}^i], \qquad p_i^1 = \Pr[R(x_i) \in U_1\backslash T_{\delta_i}^i]$$
where $T_{\delta_i}^i$ represents the failure set on $x_i$ with $\delta_i$.
\begin{figure}
\setlength{\abovecaptionskip}{5pt}%
\setlength{\belowcaptionskip}{5pt}%
  \centering
  \subfigure[Examples of $p_i^0$ v.s. $p_i^1$, $x_i$ varies within ${[}x_1^0, x_1^1{]}$. \textcolor{black}{The fluctuation on orange line suggests the different confounding patterns under personalized privacy.} ]{
  \includegraphics[width=0.4\linewidth, trim=5 5 5 5,clip]{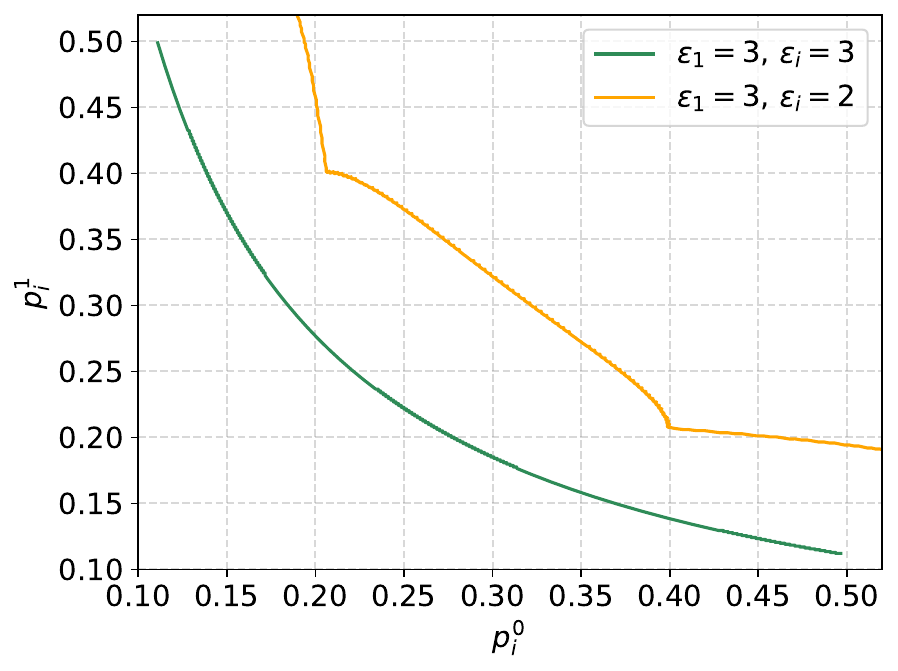}
  \label{p0_p1}
  }
 \hspace{0.5in}
  \subfigure[Fix $ \epsilon_1=3$, $p_i$ changes as $\epsilon_i$ varies. $p_i$ is highest at around $\epsilon_1=1$, as the testing error maximizes when $R(x_i)$ is almost covered by $R(x_1^0)$ and $R(x_1^1)$.]{
  \includegraphics[width=0.4\linewidth, trim=5 5 5 5,clip]{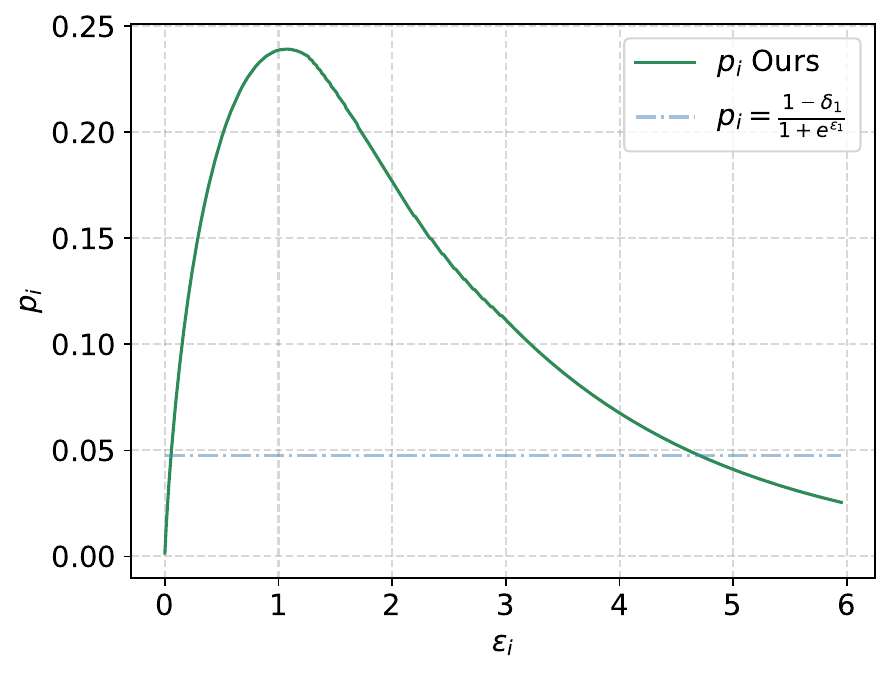}
  \label{pics_p1}
  }
  \caption{Confounding effect $p$ under personalized privacy.}
\end{figure}

We observe that $p_i^0$ and $p_i^1$ change as $x_i$ changes (Cf. Fig. \ref{p0_p1}). As $x_1^0 \leq x_i \leq x_1^1$, the worst-case happens when $x_i=x_1^0$ (or $x_i=x_1^1$). Considering the fact that privacy is breached at the weakest spot, we adopt minimal $p_i$ to describe the confounding effect of $R(x_i)$. Hence Eq.\eqref{eq_FV_3} is rewritten with $p_i = \min(p_i^0, p_i^1)$:
\begin{align}
\forall i \in [2, n], R(x_i) = p_{i}R(x_1^0) + p_{i}R(x_1^1) + (1-2p_{i})LO(x_i). 
\end{align}


\subsection{Privacy Amplification with $f$-DP}
In this section, we achieve a tighter privacy bound of shuffle model under $(\epsilon_i,\delta_i)$-PLDP with $f$-DP. 

After deriving $p$, the clones of $R(x_1^b)$ by shuffling are generated. Based on \cite{feldman2022hiding}, 
the overall distributions of number of clones on $D_0$ and $D_1$ are denoted as $P$ and $Q$, with $w=p_{1}$,
$$P = (1-w)P_0 + wQ_0 \quad \text{and}\quad Q = (1-w)Q_0 + wP_0.$$ 
where $P_0 \sim (A+1, C-A)$, $Q_0 \sim (A, C-A+1)$ with $A \sim Bin(C, 1/2)$. Considering $p_i$ varies under PLDP, we have $C_i \sim Bern(2p_{i})$, $C=\sum_{i=1}^{n-1}C_i$. 
Following the idea in \cite{wang2024unified} that mixed distributions are more indistinguishable when indices are unknown, the lower bound 
of trade-off function of overall distribution 
could be derived by establishing trade-off function on sub-distributions for each possible situation with certain weights.
Specifically, $P_0$ is the mixture of $\{(A_i+1, i-A_i)\}_{i=0}^{n-1}$ with weights $w_i^0=\Pr[C=i]$, $Q_0$ is the mixture of $\{(A_i, i-A_i+1)\}_{i=0}^{n-1}$ with the same $w_i^0$ and $A_i \sim Bin(i, 1/2)$. Let $f_i$, $F_i$ be the probability mass function and distribution function of~ $Bin(i, 1/2)$ respectively. 
By Lemma 3.1 in \cite{wang2024unified}, we achieve trade-off function $f_s$ under both pure-PLDP (let $\delta_i=0$) and approximate-PLDP settings ($\delta_i>0$). 

\begin{theorem}[Trade-off function] The trade-off function of shuffling process is defined as $f_s(\alpha(t))$, for $t\geq 0$, each  
$ \alpha(t) = \sum_{i=0}^{n-1} w_i^0 F_i(i-\frac{i+1}{t+1}) \in [0,1]$.
The function $f_s$ at $\alpha(t)$ is
    $$f_s(\alpha(t)) =\!(1\!-\!\delta_1) (2w(1\!-\!\alpha(t)) \!+\! (1\!-\!2w)\sum\nolimits_{i=0}^{n-1}w_i^0 F_i)  \!+\! \delta_1(1\!-\!\alpha(t)) $$
where $F_i$ is the abbreviation of $F_i(i+1-\frac{i+1}{t+1})$.
\end{theorem}
Then we convert it to DP based on primal-dual perspective \cite{dong2022gaussian}.
\begin{theorem}[Enhanced Privacy Bound]
The shuffling process (with randomizer, shuffler, and analyzer) $R\circ S \circ A$ is $(\epsilon, \delta_{s}(\epsilon))-DP$ for any $\epsilon>0$ with
    \begin{align}
        \delta_{s}(\epsilon) = (-e^\epsilon+ 
        (1-\delta_1)2w + \delta_1)[\sum\nolimits_{i=1}^{n-1} w_i^0 F_i(i-\frac{i+1}{t_\epsilon+1})]
        + (1-\delta_1)(1-2w)[\sum\nolimits_{i=1}^{n-1} w_i^0 F_i(i+1-\frac{i+1}{t_\epsilon+1})]
    \end{align}
    where $t_\epsilon=\inf\{t:(1-\delta_1)(-2w+(1-2w)l(t)) - \delta_1 \geq-e^\epsilon\}$ , $w=p_{1}$, $l(t)=-{\sum_{i=1}^{n-1} w_1^0 f_i(\lfloor i+1-\frac{i+1}{t+1} \rfloor)}/{\sum_{i=1}^{n-1} w_1^0 f_i(\lfloor i-\frac{i+1}{t+1} \rfloor)}$.
\end{theorem}
Noticing that the analysis on $p$ and trade-off function is related to $\epsilon_1$, $\delta_1$, which can be any of the personalized privacy parameters. 
Here we bound the worst case: user 1 with $x_1^b$ adopts weakest privacy budget, $\epsilon_1=\max(\epsilon_i)$. (Considering $\delta_i$ is negligible in usual setting, $\delta_1$ is the corresponding parameter).
\section{Experiment Results}
We show the privacy bound with various personalized privacy settings, and the different number of users.

\begin{table}
  \centering
  \begin{tabular}{llll}
    \toprule
    Name     & $\epsilon^l=\{\epsilon^l_i\}_{i\in[n]}$ & $\delta^l=\{\delta^l_i\}_{i\in[n]}$ & clip range   \\
    \midrule
    Uniform1 & $\mathcal{U}(0.05, 1)$ & $0,10^{-10}$ & $[0.05, 1]$ \\
    Gauss1 & $\mathcal{N}(0.8, 0.5)$ & $0,10^{-10}$ & $[0.05, 1]$ \\
    Uniform2 & $\mathcal{U}(0.5, 2)$ & $0,10^{-10}$ & $[0.5, 2]$ \\
    Gauss2 & $\mathcal{N}(1.5, 0.5)$ & $0,10^{-10}$ & $[0.5, 2]$ \\
    \bottomrule
  \end{tabular}

  \caption{PLDP privacy parameters ${\epsilon^l,\delta^l}$. For pure PLDP set $\delta^l$ as $0$. $\mathcal{U}$, $\mathcal{N}$ represent Uniform and Gaussian Distribution respectively. The $\delta^s$ after shuffling is $10^{-5}$.}
  \label{table_eps_dist}
\end{table} 

\textbf{Experiment Setting.} We evaluate  several PLDP parameter settings as Tab.\ref{table_eps_dist}. Baselines include: for pure DP, BBGN\cite{balle2019privacy}, FV\cite{feldman2023stronger}, CCC\cite{chen2024generalized}, LZX\cite{liu2023echo}; for approximate DP, FV\cite{feldman2023stronger}, CCC\cite{chen2024generalized}. Notice that BBGN and FV lack the analysis on personalized privacy, only the approximate bound is demonstrated by using $\max(\epsilon_i)$ for all data points. We set the same $\delta^l$ for all users  for convenience, as FV is easy to be unbound with large $\delta^l$. For our bound, we select Laplace Mechanism and Gaussian Mechanism for evaluating pure and approximate-PLDP 
respectively. In practical application, 
our analysis allows 
any personalized $\delta_i$ and local randomizers.

\begin{figure}[!t]
\setlength{\abovecaptionskip}{5pt}%
\setlength{\belowcaptionskip}{5pt}%
  \centering
  \subfigure[$\epsilon^l_i$-PLDP where $\epsilon^l_i \in {[} 0.05,1{]} $.]{
  \includegraphics[width=0.4\linewidth, trim=0 0 5 37,clip]{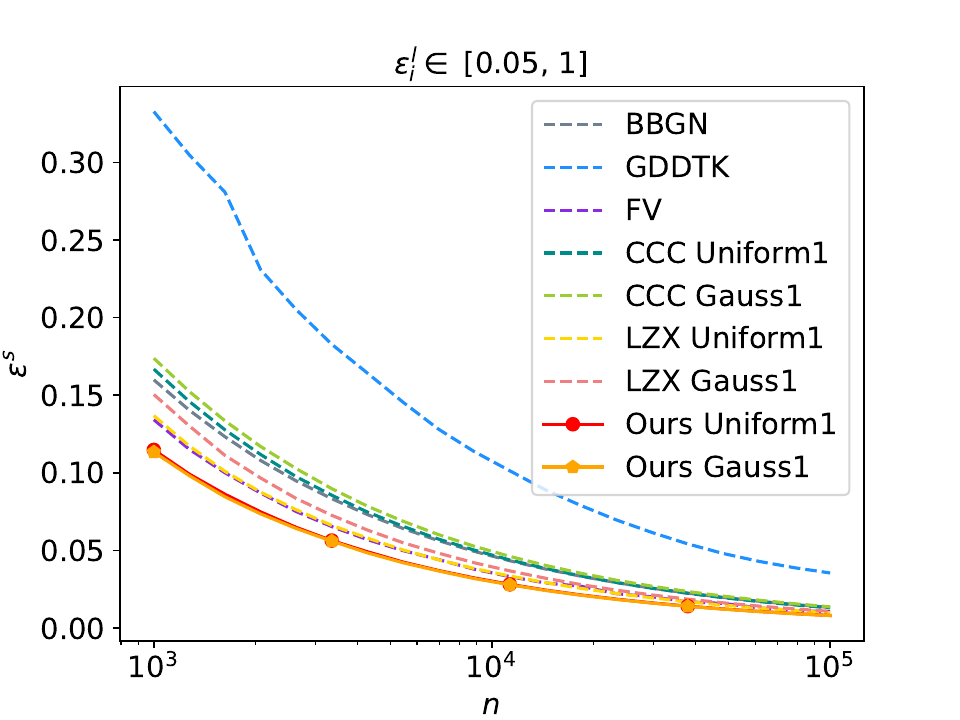}
  \label{pics_amplify_pure}
  }
 \hspace{0.1in}
  \subfigure[$(\epsilon^l_i, \delta^l_i)$-PLDP, where $\epsilon^l_i \in {[}0.05,1{]}$.]{
  \includegraphics[width=0.4\linewidth, trim=0 0 5 37,clip]{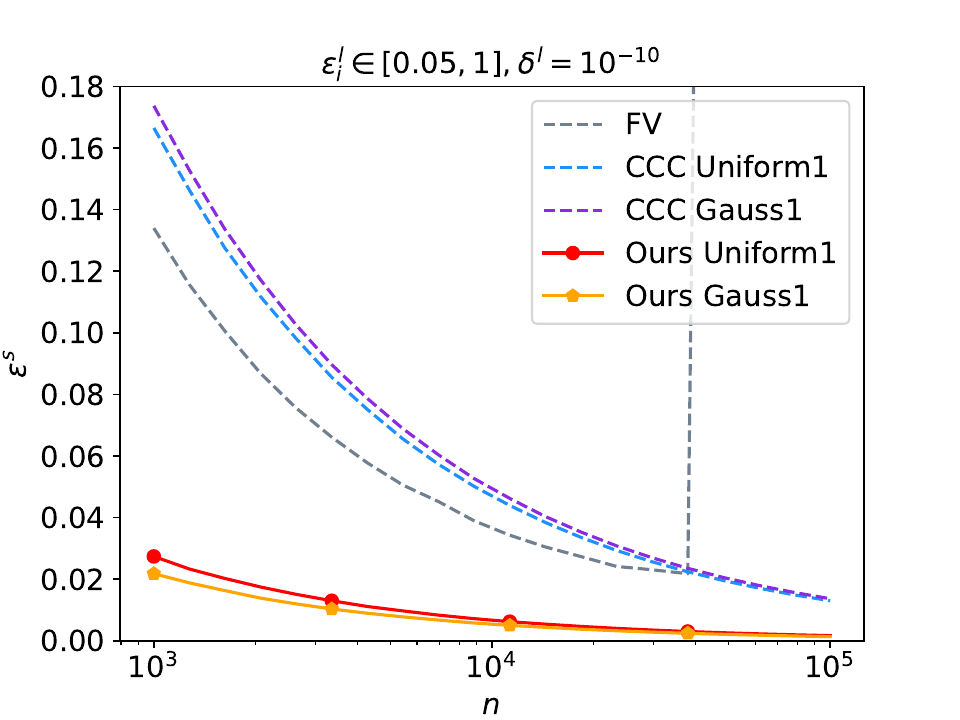}
  \label{pics_amplify_approx}
  }
  \subfigure[$\epsilon^l_i$-PLDP where $\epsilon^l_i \in {[} 0.5,2{]} $.]{
  \includegraphics[width=0.4\linewidth, trim=0 0 5 37,clip]{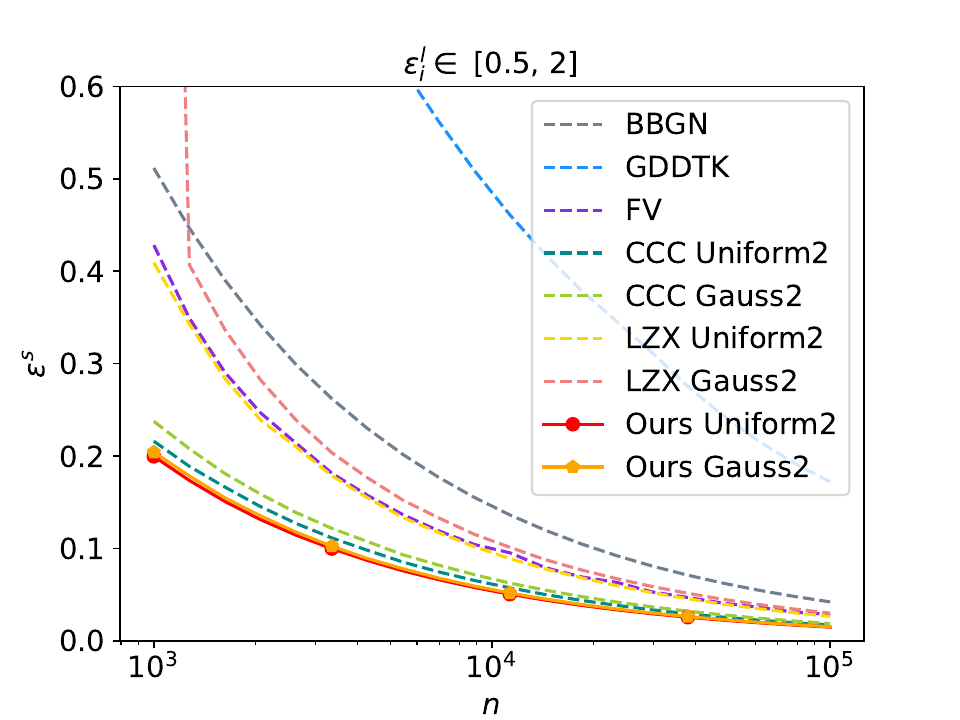}
  \label{pics_amplify_pure}
  }
 \hspace{0.1in}
  \subfigure[$(\epsilon^l_i, \delta^l_i)$-PLDP, where $\epsilon^l_i \in {[}0.5,2{]}$.]{
  \includegraphics[width=0.4\linewidth, trim=0 0 5 37,clip]{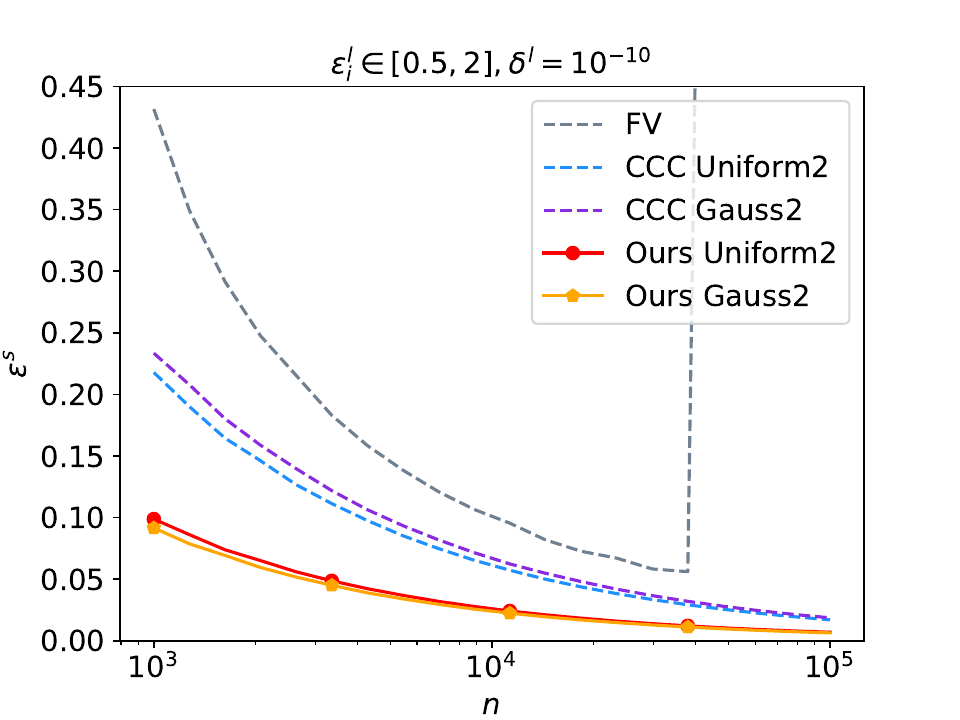}
  \label{pics_amplify_approx}
  }
  \caption{Privacy bounds with various number of data points and privacy parameters, for pure- and approximate-PLDP.}
\label{pics_bound_eps}
\end{figure}

\textbf{Privacy Amplification with fixed $\delta^s$.}
Fig.\ref{pics_bound_eps} provides the numerical evaluations for privacy amplification effect with various PLDP settings and the number of users. (1) Our bound achieves the strongest privacy amplification effect. The results come from a precise $p$ with hypothesis testing on concrete mechanism, and sharp bound with $f$-DP.
(2) Compared to pure-PLDP, bound on approximate-PLDP is tighter. It is reasonable from two aspects: first, Gaussian Mechanism is much more noisy (larger variance) than Laplace Mechanism under the same $\epsilon$. Hence the confounding effect $p$ is larger on approximate-PLDP; second, $f$-DP precisely characterizes the Gaussian distribution, hence the bound is tighter on approximate-PLDP.

\textbf{Privacy Amplification with fixed $\epsilon^s$.}
Tab.\ref{tab_delta} presents values of $\delta^s$ after shuffling with different fixed $\epsilon^s$ values. Due to limited space we only show the result of approximate-PLDP, the performance on pure-PLDP is similar. Notably, under the same $\epsilon^s$, our bound on $\delta$ is significantly smaller than baselines. 

\begin{table}
  \centering
  \begin{tabular}{llllll}
    \toprule
    $\epsilon^s$     & 0.01 & 0.03 & 0.05 & 0.08 & 0.1    \\
    \midrule
    $\delta^s$ \cite{feldman2023stronger} & $0.0083$ & $0.0029$ & $0.0008$ & $6\times 10^{-5}$ & $1\times 10^{-5}$ \\
    $\delta^s$ \cite{chen2024generalized} & $0.0042$ & $0.0007$ & $5\times 10^{-5}$ & $3\times 10^{-7}$ & $3\times 10^{-9}$\\
    $\delta^s$~(Ours) & $0.0007$ & $2\times 10^{-6}$ & $3\times 10^{-10}$ & $1\times 10^{-18}$ & $1\times 10^{-25}$ \\
    \bottomrule
\end{tabular}
  \caption{$\delta^s$ after shuffling comparison under $(\epsilon^l, \delta^l)$-PLDP with Uniform2, $n=10000$.}
\label{tab_delta}
\end{table} 


\section{Conclusion}
This work achieves a refined privacy bound on shuffle model for both pure- and approximate-PLDP. To tighten the bound, we provide a full analysis on confounding effect of perturbed individual data and the overall distributions. Our bound on $\epsilon$ is up to $5$ times smaller than SOTAs. 

\section{Acknowledgement}
This research has been funded in part by the National Key Research \& Develop Plan 2023YFB4503600; National Natural Science Foundation of China~(NSFC) U23A20299, 62072460, 62172424, 62276270, 62322214;  National Science Foundation~(NSF) CNS-2124104, CNS-2125530, IIS-2302968; National Institute of Health~(NIH) R01LM013712, R01ES033241. 
The corresponding author is Hong Chen.
\bibliographystyle{plain}
\balance
\bibliography{main}

\begin{thebibliography}{10}

\bibitem{balle2019privacy}
Borja Balle, James Bell, Adri{\`a} Gasc{\'o}n, and Kobbi Nissim.
\newblock The privacy blanket of the shuffle model.
\newblock In {\em Annual International Cryptology Conference}, pages 638--667. Springer, 2019.

\bibitem{bittau2017prochlo}
Andrea Bittau, {\'U}lfar Erlingsson, Petros Maniatis, Ilya Mironov, Ananth Raghunathan, David Lie, Mitch Rudominer, Ushasree Kode, Julien Tinnes, and Bernhard Seefeld.
\newblock Prochlo: Strong privacy for analytics in the crowd.
\newblock In {\em Proceedings of the 26th symposium on operating systems principles}, pages 441--459, 2017.

\bibitem{chen2024generalized}
E~Chen, Yang Cao, and Yifei Ge.
\newblock A generalized shuffle framework for privacy amplification: Strengthening privacy guarantees and enhancing utility.
\newblock In {\em Proceedings of the AAAI Conference on Artificial Intelligence}, volume~38, pages 11267--11275, 2024.

\bibitem{chen2024privacy}
Wei-Ning Chen, Dan Song, Ayfer Ozgur, and Peter Kairouz.
\newblock Privacy amplification via compression: Achieving the optimal privacy-accuracy-communication trade-off in distributed mean estimation.
\newblock {\em Advances in Neural Information Processing Systems}, 36, 2024.

\bibitem{cormode2018privacy}
Graham Cormode, Somesh Jha, Tejas Kulkarni, Ninghui Li, Divesh Srivastava, and Tianhao Wang.
\newblock Privacy at scale: Local differential privacy in practice.
\newblock In {\em Proceedings of the 2018 International Conference on Management of Data}, pages 1655--1658, 2018.

\bibitem{dong2022gaussian}
Jinshuo Dong, Aaron Roth, and Weijie~J Su.
\newblock Gaussian differential privacy.
\newblock {\em Journal of the Royal Statistical Society Series B: Statistical Methodology}, 84(1):3--37, 2022.

\bibitem{dwork2006calibrating}
Cynthia Dwork, Frank McSherry, Kobbi Nissim, and Adam Smith.
\newblock Calibrating noise to sensitivity in private data analysis.
\newblock In {\em Theory of Cryptography: Third Theory of Cryptography Conference, TCC 2006, New York, NY, USA, March 4-7, 2006. Proceedings 3}, pages 265--284. Springer, 2006.

\bibitem{dwork2014algorithmic}
Cynthia Dwork, Aaron Roth, et~al.
\newblock The algorithmic foundations of differential privacy.
\newblock {\em Foundations and Trends{\textregistered} in Theoretical Computer Science}, 9(3--4):211--407, 2014.

\bibitem{erlingsson2019amplification}
{\'U}lfar Erlingsson, Vitaly Feldman, Ilya Mironov, Ananth Raghunathan, Kunal Talwar, and Abhradeep Thakurta.
\newblock Amplification by shuffling: From local to central differential privacy via anonymity.
\newblock In {\em Proceedings of the Thirtieth Annual ACM-SIAM Symposium on Discrete Algorithms}, pages 2468--2479. SIAM, 2019.

\bibitem{erlingsson2014rappor}
{\'U}lfar Erlingsson, Vasyl Pihur, and Aleksandra Korolova.
\newblock Rappor: Randomized aggregatable privacy-preserving ordinal response.
\newblock In {\em Proceedings of the 2014 ACM SIGSAC conference on computer and communications security}, pages 1054--1067, 2014.

\bibitem{feldman2022hiding}
Vitaly Feldman, Audra McMillan, and Kunal Talwar.
\newblock Hiding among the clones: A simple and nearly optimal analysis of privacy amplification by shuffling.
\newblock In {\em 2021 IEEE 62nd Annual Symposium on Foundations of Computer Science (FOCS)}, pages 954--964. IEEE, 2022.

\bibitem{feldman2023stronger}
Vitaly Feldman, Audra McMillan, and Kunal Talwar.
\newblock Stronger privacy amplification by shuffling for r{\'e}nyi and approximate differential privacy.
\newblock In {\em Proceedings of the 2023 Annual ACM-SIAM Symposium on Discrete Algorithms (SODA)}, pages 4966--4981. SIAM, 2023.

\bibitem{girgis2021shuffled}
Antonious Girgis, Deepesh Data, Suhas Diggavi, Peter Kairouz, and Ananda~Theertha Suresh.
\newblock Shuffled model of differential privacy in federated learning.
\newblock In {\em International Conference on Artificial Intelligence and Statistics}, pages 2521--2529. PMLR, 2021.

\bibitem{girgis2024multi}
Antonious~M Girgis and Suhas Diggavi.
\newblock Multi-message shuffled privacy in federated learning.
\newblock {\em IEEE Journal on Selected Areas in Information Theory}, 5:12--27, 2024.

\bibitem{kairouz2015composition}
Peter Kairouz, Sewoong Oh, and Pramod Viswanath.
\newblock The composition theorem for differential privacy.
\newblock In {\em International conference on machine learning}, pages 1376--1385. PMLR, 2015.

\bibitem{lehmann2006theory}
Erich~L Lehmann and George Casella.
\newblock {\em Theory of point estimation}.
\newblock Springer Science \& Business Media, 2006.

\bibitem{liu2021projected}
Junxu Liu, Jian Lou, Li~Xiong, Jinfei Liu, and Xiaofeng Meng.
\newblock Projected federated averaging with heterogeneous differential privacy.
\newblock {\em Proceedings of the VLDB Endowment}, 15(4):828--840, 2021.

\bibitem{liu2024cross}
Junxu Liu, Jian Lou, Li~Xiong, Jinfei Liu, and Xiaofeng Meng.
\newblock Cross-silo federated learning with record-level personalized differential privacy.
\newblock {\em arXiv preprint arXiv:2401.16251}, 2024.

\bibitem{liu2023personalized}
Junxu Liu, Jian Lou, Li~Xiong, and Xiaofeng Meng.
\newblock Personalized differentially private federated learning without exposing privacy budgets.
\newblock In {\em Proceedings of the 32nd ACM International Conference on Information and Knowledge Management}, pages 4140--4144, 2023.

\bibitem{liu2021privacy}
Yixuan Liu, Hong Chen, Yuhan Liu, and Cuiping Li.
\newblock Privacy-preserving techniques in federated learning.
\newblock {\em Journal of Software}, 33(3):1057--1092, 2021.

\bibitem{liu2023echo}
Yixuan Liu, Suyun Zhao, Li~Xiong, Yuhan Liu, and Hong Chen.
\newblock Echo of neighbors: privacy amplification for personalized private federated learning with shuffle model.
\newblock In {\em Proceedings of the AAAI Conference on Artificial Intelligence}, volume~37, pages 11865--11872, 2023.

\bibitem{liu2024edge}
Yuhan Liu, Tianhao Wang, Yixuan Liu, Hong Chen, and Cuiping Li.
\newblock Edge-protected triangle count estimation under relationship local differential privacy.
\newblock {\em IEEE Transactions on Knowledge and Data Engineering}, 2024.

\bibitem{liu2022collecting}
Yuhan Liu, Suyun Zhao, Yixuan Liu, Dan Zhao, Hong Chen, and Cuiping Li.
\newblock Collecting triangle counts with edge relationship local differential privacy.
\newblock In {\em 2022 IEEE 38th International Conference on Data Engineering (ICDE)}, pages 2008--2020. IEEE, 2022.

\bibitem{scott2022aggregation}
Mary Scott, Graham Cormode, and Carsten Maple.
\newblock Aggregation and transformation of vector-valued messages in the shuffle model of differential privacy.
\newblock {\em IEEE Transactions on Information Forensics and Security}, 17:612--627, 2022.

\bibitem{wang2024unified}
Chendi Wang, Buxin Su, Jiayuan Ye, Reza Shokri, and Weijie Su.
\newblock Unified enhancement of privacy bounds for mixture mechanisms via $ f $-differential privacy.
\newblock {\em Advances in Neural Information Processing Systems}, 36, 2024.

\bibitem{wang2019collecting}
Ning Wang, Xiaokui Xiao, Yin Yang, Jun Zhao, Siu~Cheung Hui, Hyejin Shin, Junbum Shin, and Ge~Yu.
\newblock Collecting and analyzing multidimensional data with local differential privacy.
\newblock In {\em 2019 IEEE 35th International Conference on Data Engineering (ICDE)}, pages 638--649. IEEE, 2019.

\end{thebibliography}



\end{document}